\documentclass[aps,superscriptaddress,altaffilletter]{revtex4}%
\usepackage{amssymb}
\usepackage{amsfonts}
\usepackage{amsmath}
\usepackage{graphicx}%
\providecommand{\U}[1]{\protect\rule{.1in}{.1in}}

\begin{document}
\title[Phase transitions and pseudogap]{Pseudogap and preformed pairs in the imbalanced Fermi gas in two dimensions}
\author{S.N. Klimin}
\altaffiliation[On leave of absence from:]{Department of Theoretical Physics, State University of Moldova, str. A.
Mateevici 60, MD-2009 Kishinev, Republic of Moldova.}

\affiliation{Theorie van Kwantumsystemen en Complexe Systemen, Universiteit Antwerpen,
Universiteitsplein 1, B-2610 Antwerpen, Belgium}
\author{J. Tempere}
\affiliation{Theorie van Kwantumsystemen en Complexe Systemen, Universiteit Antwerpen,
Universiteitsplein 1, B-2610 Antwerpen, Belgium}
\affiliation{Lyman Laboratory of Physics, Harvard University, Cambridge, MA 02138, USA}
\author{J.T. Devreese}
\altaffiliation[Also at:]{Technische Universiteit Eindhoven, P. B. 513, 5600 MB Eindhoven, The Netherlands.}

\affiliation{Theorie van Kwantumsystemen en Complexe Systemen, Universiteit Antwerpen,
Universiteitsplein 1, B-2610 Antwerpen, Belgium}

\begin{abstract}
The physics of the pseudogap state is intimately linked with the pairing
mechanism that gives rise to superfluidity in quantum gases and to
superconductivity in high-$T_{c}$ cuprates, and therefore, both in quantum
gases and superconductors, the pseudogap state and preformed pairs have been
under intensive experimental scrutiny. Here, we develop a path integral
treatment that provides a divergence-free description of the paired state in
two-dimensional Fermi gases. Within this formalism, we derive the pseudogap
temperature and the pair fluctuation spectral function, and compare these
results with the recent experimental measument of the pairing in the
two-dimensional Fermi gas. The removal of the infrared divergence in the
number equations is shown both numerically and analytically, through a study
of the long-wavelength and low-energy limit of the pair fluctuation density.
Besides the pseudogap temperature, also the pair formation temperature and the
critical temperature for superfluidity are derived. The latter corresponds to
the Berezinski-Kosterlitz-Thouless (BKT) temperature. The pseudogap
temperature, which coincides with the pair formation temperature in mean
field, is found to be suppressed with respect to the pair formation
temperature by fluctuations. This suppression is strongest for large binding
energies of the pairs. Finally, we investigate how the pair formation
temperature, the pseudogap temperature and the BKT temperature behave as a
function of both binding energy and imbalance between the pairing partners in
the Fermi gas. This allows to set up phase diagrams for the two-dimensional
Fermi gas, in which the superfluid phase, the phase-fluctuating
quasicondensate, and the normal state can be identified.

\end{abstract}
\date{\today}
\maketitle

\bigskip

\section{Introduction}

Ultracold atomic gases are increasingly used as quantum simulators to probe
many-body physics \cite{DalibardReview}. Recent efforts have focused in
particular on understanding superfluidity and Cooper pairing in interacting
Fermi systems, including \emph{i.a.} the effects of varying the interaction
strength, introducing population imbalance, and reducing the dimensionality.
When these fermionic superfluids are described in the path integral formalism,
the thermodynamic potential $\Omega(T,\mu_{\uparrow},\mu_{\downarrow})$ of the
interacting Fermi gas (as a function of temperature $T$ and chemical
potentials $\mu_{\uparrow}$,$\mu_{\downarrow}$ of spin-up and spin-down
components) is rewritten as a functional integral over a bosonic field
$\Delta_{\mathbf{x},\tau}$ that represents the field of the pairs. This field
is introduced through the Hubbard-Stratonovich transformation that allows the
exact elimination of the fermionic degrees of freedom and results in an action
functional for the bosonic field \cite{deMelo1993}. At first sight, one has
only succeeded in rewriting an unsolvable functional integral over fermionic
fields by an equally unsolvable functional integral over the bosonic fields.
However, the bosonic field lends itself to an obvious simplification when one
intuits that a (uniform) Bose-Einstein condensation of pairs is present. In
that case, one can surmise that $\Delta_{\mathbf{x},\tau}\approx\Delta$, i.e.
all pairs are in the zero-momentum state so the field is a constant in real
space. The functional integral can then replaced by its saddle-point value,
substituting $\Delta_{\mathbf{x},\tau}\approx\Delta$ and dropping the
integrations. The optimal value of $\Delta$ is found by extremizing the action
or, equivalently, by minimizing the thermodynamic potential $\Omega_{sp}%
(T,\mu_{\uparrow},\mu_{\downarrow};\Delta)$ with respect to the saddle point.
When $\Delta=0$, the (non-interacting) normal Fermi gas is obtained, when
$\Delta\neq0$, the saddle-point approximation bears out a Bose-Einstein
condensate of pairs.

In two-dimensional superfluid systems, the interpretation of the bosonic field
is more subtle. To be precise, the condition for Bose-Einstein condensation
(BEC) has been defined by Penrose, Onsager \cite{PO} and Yang \cite{Yang} as
the presence of off-diagonal long range order, i.e. $\lim_{\mathbf{x}%
-\mathbf{x}^{\prime}\rightarrow\infty}\left\langle \Delta_{\mathbf{x},\tau
}\Delta_{\mathbf{x}^{\prime},\tau}\right\rangle \neq0$. At the level of the
saddle point, $\Delta_{\mathbf{x},\tau}\approx\Delta$, it is clear that a
nonzero $\Delta$ implies BEC. However, as Mermin,Wagner \cite{MW} and
Hohenberg \cite{Hohenberg} pointed out, in the two-dimensional system
fluctuations play a crucial role: they will prohibit off-diagonal long range
order in uniform systems. These fluctuations around the saddle point are
commonly taken into account through the Bogoliubov shift $\Delta
_{\mathbf{x},\tau}=\Delta+\phi_{\mathbf{x},\tau}$ , whereafter $\phi
_{\mathbf{x},\tau}$ is treated as a small fluctuation so that only terms up to
second order in $\phi_{\mathbf{x},\tau}$ are retained in the action
functional. Then, for a given $\Delta$, the thermodynamic potential
$\Omega(T,\mu_{\uparrow},\mu_{\downarrow};\Delta)$ is expressed as a
functional integral over the fluctuation fields, with a quadratic action. The
quadratic functional integral can be performed, and we obtain a fluctuation
correction $\Omega_{fl}(T,\mu_{\uparrow},\mu_{\downarrow};\Delta)=\Omega
(T,\mu_{\uparrow},\mu_{\downarrow};\Delta)-\Omega_{sp}(T,\mu_{\uparrow}%
,\mu_{\downarrow};\Delta)$ to the thermodynamic potential. The fluctuation
fields need not be written down as complex fields $\phi_{\mathbf{x},\tau}$
resulting from the Bogoliubov shift: equivalent results are obtained by
introducing (real) amplitude and phase fluctuation fields through
$\Delta_{\mathbf{x},\tau}\approx\Delta\left(  1+\delta_{\mathbf{x},\tau
}\right)  e^{i\theta_{\mathbf{x},\tau}}$.

The culprit suppressing off-diagonal long range order in 2D is precisely the
phase fluctuation field $e^{i\theta_{\mathbf{x},\tau}}$. Indeed, Mermin and
Wagner show that $\left\langle \Delta_{\mathbf{x},\tau}\Delta_{\mathbf{x}%
^{\prime},\tau}\right\rangle \approx\left\langle \Delta^{2}e^{i\theta
_{\mathbf{x},\tau}}\right\rangle \rightarrow0$ due to the long-wavelength
behavior of $\theta_{\mathbf{x},\tau}$, the relative phase. According to the
Penrose-Onsager-Yang criterion this means that Bose-Einstein condensation does
not occur. However, we can identify other interesting phases from a study of
the bosonic pair field $\Delta_{\mathbf{x},\tau}$. Firstly, $\left\langle
e^{i\theta_{\mathbf{x},\tau}}\right\rangle \rightarrow0$ does not imply that
$\Delta=0,$ as noted by Kagan in his study of quasicondensation \cite{Kagan}.
We can identify $\Delta\neq0$ with the presence of pairing, and search for a
transition temperature $T_{c}^{\ast}$ for pair formation separating the
$\Delta=0$ phase from the $\Delta\neq0$ low temperature phase. Second,
although BEC is suppressed, superfluidity can still be present in the two
dimensional system below the Berezinski-Kosterlitz-Thouless
\cite{Berezinskii,KT} (BKT) temperature $T_{BKT}$. The order parameter for
superfluidity is $\rho_{s}$, the superfluid density, defined as the phase
stiffness and calculated as the prefactor of the $\left(  \mathbf{\nabla
}\theta_{\mathbf{x},\tau}\right)  ^{2}$ term in the Lagrangian for the phase
field, as explained in more detail below. Kosterlitz and Thouless \cite{KT}
described a mechanism whereby phase stiffness can be lost, namely through the
appearance and unbinding of vortex-antivortex pairs that start to proliferate
at $T_{BKT}$ and scramble the phase field. This mechanism was observed
experimentally in an 2D atomic Bose gas by Dalibard and co-workers
\cite{DalibardBKT}.

When the bosons under consideration are composite particles, such as Cooper
pairs, a third relevant temperature can be identified, related to the density
of states of excitations, or equivalently the spectral function for the
fluctuations. As we will show below, the fluctuation terms in the density can
be expressed through a spectral function $g(\mathbf{q},\omega)$ describing the
contribution of fluctuations with a given wave number $\mathbf{q}$ and
momentum $\omega$. In the Nozi\`{e}res and\ Schmitt-Rink (NSR) formalism
\cite{NSR} for the two dimensional system \cite{NSR2D}, the integral over the
spectral function is divergent, invalidating the number equations
$\partial\Omega/\partial\mu_{\sigma}=n_{\sigma}$, $\sigma=\uparrow,\downarrow
$. We show that this divergency is absent when we apply the formalism of Hu,
Liu and Drummond \cite{Hu,Drum,Drum2}, which these authors dubbed the Gaussian
Pair Fluctuation (GPF) approach, to the two dimensional case. As we show below
(section \ref{gpf}), the GPF approach allows to set up and simultaneously
solve the gap and number equations also in the two-dimensional case. This
allows us to derive results for $\Delta,\rho_{s}$ that take into account
fluctuations (both phase fluctuations and amplitude fluctuations). The
resulting fluctuation spectra can then be used to obtain the
finite-temperature thermodynamics of the two-dimensional Fermi superfluid,
following the approach of Salasnich for the three-dimensional case
\cite{Salasnich}.

In studying the fluctuation spectra, we find that for temperatures above a
critical temperature $T_{p}$, the fluctuation spectral function $g(\mathbf{q}%
,\omega)$ becomes negative at long wavelengths ($q<q_{c}$). This happens at a
temperature above $T_{BKT}$ and (obviously) below $T_{c}^{\ast}$. We interpret
this temperature $T_{p}$ as the pairing temperature at which the pseudogap is
open, inspired by the recent experiments \cite{Feld,Sommer} which investigated
pairing in ultracold 2D atomic Fermi gases. For the ultracold atomic gases in
3D, the pseudogap state above the critical temperature is a subject of an
intense study, both experimental and theoretical
\cite{VDM,Gaebler,Perali,Palestrini,Magierski,Tsuchiya}, and the similarity
with the pseudogap physics in superconductors has not gone unnoticed
\cite{LevinRPP}. In the experiment \cite{Feld}, the spectral function for
excitations of the Fermi gas is determined through momentum-resolved
photoemission spectroscopy. In Ref. \cite{Sommer}, the momentum-integrated
photoemission spectra are measured, and the evolution of fermion pairing was
followed from three to two dimensions by varying the strength of the confining
optical lattice. Both experiments reveal a non-zero pairing gap. While the
experiments in \cite{Feld} were interpreted to reveal the pseudogap, i.e.
pairing in a non-superfluid state, superfluidity itself has not been observed
in 2D yet. Therefore, the existence of the pseudo-gap regime is not
experimentally settled until superfluidity itself is observed at temperatures
lower than the temperature for pair formation.

We compare the fluctuation spectral functions derived from our microscopic
(GPF-based) theory to the measured spectral functions for excitations, and we
also compare the measured pseudogap temperatures with the calculated $T_{p}$
as a function of the interaction strength. We find that fluctuations indeed
greatly lower the temperature range of existence of the pseudogap phase,
especially in the strong-coupling regime. Consequently, in order to obtain a
complete phase diagram for the Fermi gas in 2D, we must consider each phase
taking in account fluctuations. To the best of our knowledge, this problem has
hitherto not yet been satisfactorily solved for the case of fermions in 2D
because of the aforesaid divergence of the density due to fluctuations at
finite temperatures. Here, as mentioned above, we tackle the problem by
correcting the NSR approach using the Gaussian Pair Fluctuation theory (GPF)
proposed by Hu, Liu and Drummond \cite{Hu,Drum,Drum2} for the three
dimensional Fermi gas. Moreover, we extend the results to the case of imbalance.

\bigskip

The paper is organized as follows. In Sec. \ref{formalism}, we present the
divergence-free method for the self-consistent calculation of thermodynamic
parameters of interacting imbalanced fermions in 2D taking into account both
amplitude and phase fluctuations. In Sec. \ref{expansion}, density
distribution functions for an imbalanced 2D Fermi gas are investigated. In
Sec. \ref{phasediagrams}, we discuss finite-temperature phase diagrams for the
imbalanced Fermi gas in 2D. In Sec. \ref{Experiment}, the theory is applied to
the interpretation of the experiment on pairing of cold atoms in 2D. The
discussion is followed by conclusions, Sec. \ref{conclusions}.

Before going forward in the next section with presenting the functional
integral approach in the GPF framework, it is useful to note that the GPF
approach that we follow here is not the only way to avoid the divergence
problem that occurs in the NSR description for the Fermi gas in two
dimensions. The NSR scheme and its modifications are related to the $T$-matrix
perturbation approach, in which the effective interaction between pairs is
taken into account diagrammatically. In this context, the divergence problem
for a Fermi gas in two dimensions can be remedied by taking into account
higher orders of the $T$-matrix expansion -- via an effective interaction
between pair fluctuations \cite{Traven} . This interaction stabilizes the
superfluid phase of the 2D fermion system at very low temperatures. However,
in 2D the $T$-matrix method does not predict the universal jump in the
superfluid density \cite{Nelson1977} related to the BKT phase transition. A
correct description of the superfluid density becomes possible by explicitly
focusing on the phase fluctuations, as in the approach of Refs.
\cite{Babaev,Gus1,Gus2,Botelho2006}. Within that approach, bosonic pair field
is gauge transformed $\Delta_{\mathbf{x},\tau}e^{i\theta_{\mathbf{x},\tau}}$
and a subsequent gradient expansion of the fluctuation action is performed for
phase fluctuations assuming that phase gradients are small. This leads to a
quadratic effective action functional of the phase field $\theta$, which, as
distinct from the scheme of Ref. \cite{NSR}, has no divergence for $\Delta
\neq0$. As far as gradients of the fields are assumed to be small, the
resulting effective action is treated as a hydrodynamic action (see, e.g.,
Ref. \cite{Popov}). The gradient expansion does not contain the \emph{a
priori} assumption that fluctuations themselves are small. In this connection,
the method was categorized in Ref. \cite{Babaev} as non-perturbative. In Ref.
\cite{BKT-PRA2009}, the present authors applied the method of Refs.
\cite{Babaev,Gus1,Gus2,Botelho2006} to derive the effective hydrodynamic
action for a Fermi gas with a population imbalance. A non-perturbative
approach was also the key to develop a description free of infrared and
ultraviolet divergences for the 2D Bose gas \cite{Stoof}, that successfully
describes the crossover between the mean-field regime and the critical
fluctuation range corresponding the BKT transition \cite{PS2002}.

\section{Thermodynamic functions of the Fermi gas in 2D \label{formalism}}

\subsection{Gap equation}

We consider a gas of interacting fermions in 2D, with a contact interaction
and with $s$-wave pairing. In the ultracold regime where only $s$-wave
interactions matter, these interactions only take place between
\textquotedblleft spin-up\textquotedblright\ and \textquotedblleft
spin-down\textquotedblright\ fermions (in practice, these are usually two
different hyperfine states of an atomic species). The thermodynamic functions
of the Fermi gas are completely determined by the partition function. Here we
will focus on the thermodynamic potential $\Omega$ per unit area. The
treatment is performed within the path-integral formalism following Ref.
\cite{BKT-PRA2009}, building on the original path-integral treatment in Ref.
\cite{deMelo1993} for the case of a balanced three-dimensional Fermi gas. The
partition function is represented as the path integral over Grassmann
variables $\bar{\psi}_{\sigma}\left(  \mathbf{x},\tau\right)  ,\psi_{\sigma
}\left(  \mathbf{x},\tau\right)  $,%
\begin{equation}
\mathcal{Z}=e^{-\beta\Omega(T,\mu,\zeta)}=\int\mathcal{D}\psi_{\sigma
,\mathbf{x},\tau}\mathcal{D}\bar{\psi}_{\sigma,\mathbf{x},\tau}\exp\left(
-S\right)  .
\end{equation}
The action functional of interacting fermions is given by the integral%
\begin{align}
S  &  =\int_{0}^{\beta}d\tau\int d^{2}\mathbf{x}\sum_{\sigma=\uparrow
,\downarrow}\bar{\psi}_{\sigma,\mathbf{x},\tau}\left(  \frac{\partial
}{\partial\tau}-\nabla_{\mathbf{x}}^{2}-\mu_{\sigma}\right)  \psi
_{\sigma,\mathbf{x},\tau}\nonumber\\
&  +g\int_{0}^{\beta}d\tau\int d^{2}\mathbf{x}~\bar{\psi}_{\uparrow
,\mathbf{x},\tau}\bar{\psi}_{\downarrow,\mathbf{x},\tau}\psi_{\downarrow
,\mathbf{x},\tau}\psi_{\uparrow,\mathbf{x},\tau}, \label{act}%
\end{align}
where $g$ is the interaction strength and $\beta=1/(k_{B}T)$ is the inverse
thermal energy. We choose a system of units where $\hbar=1$, $2m=1$, and the
Fermi wave vector $k_{F}\equiv\left(  2\pi n\right)  ^{1/2}=1$ with $n$ the
total density. Here, we consider also the case when imbalance is present, i.e.
the number of spin-up and spin-down atoms are unequal: $n_{\uparrow}\neq
n_{\downarrow}.$ This in turn implies that the chemical potentials
$\mu_{\uparrow}$ and $\mu_{\downarrow}$ should be fixed separately. Rather
than contemplating the separate components, we will work with the total
density $n=n_{\uparrow}+n_{\downarrow}$ and the density difference $\delta
n=n_{\uparrow}-n_{\downarrow}$. Correspondingly, we will use the average
chemical potential $\mu=(\mu_{\uparrow}+\mu_{\downarrow})/2,$ and the chemical
potential difference $\zeta=(\mu_{\uparrow}-\mu_{\downarrow})/2$. Note that
the total density is equal to $1/(2\pi)$ in our units, so this means that we
need to solve the number equation to fix $\mu$ (in the non-interacting case,
$\mu=1$ in our units). Only with respect to the imbalance, we have a choice of
studying the free energy (and making phase diagrams as a function of $\delta
n$) or the thermodynamic potential $\Omega$ (and making phase diagrams as a
function of $\zeta$). The thermodynamic potential is linked to the free energy
by the usual Legendre transform and, as mentioned, in our formalism this
corresponds to imposing the number equations.

The strength $g$ of the contact interaction is renormalized as in Refs.
\cite{R1990,Tempere2007} using the binding energy $E_{b}$ for a two-particle
bound state, which always exists in 2D \cite{Petrov,Michiel}:%
\begin{equation}
\frac{1}{g}=\frac{1}{8\pi}\left(  \ln\frac{E_{b}}{E}+i\pi\right)  -\int
\frac{d^{2}\mathbf{k}}{\left(  2\pi\right)  ^{2}}\frac{1}{2k^{2}-E+i\delta}.
\label{renorm}%
\end{equation}
with $\delta$ a positive infinitesimal number. The BCS regime corresponds to
$E_{b}/E_{F}\ll1$, whereas the BEC regime corresponds to the opposite ratio
$E_{b}/E_{F}\gg1$. Similarly to Ref. \cite{deMelo1993}, we introduce the pair
field $\Delta_{\mathbf{x},\tau}$ and perform the Hubbard-Stratonovich
transformation, which results in a fermion-boson action quadratic in fermion
variables. After integrating out the fermion variables, the following
effective bosonic action is obtained as a functional of the
Hubbard-Stratonovich pair field $\Delta_{\mathbf{x},\tau}$:%
\begin{equation}
S_{eff}=-\operatorname*{tr}\left[  \ln\left(  -\mathbb{G}^{-1}\right)
\right]  -\int_{0}^{\beta}d\tau\int d^{2}\mathbf{x}\frac{\bar{\Delta
}_{\mathbf{x},\tau}\Delta_{\mathbf{x},\tau}}{g}, \label{Seff}%
\end{equation}
where $\mathbb{G}^{-1}$ is the inverse of the Nambu propagator%
\begin{equation}
-\mathbb{G}^{-1}=\sigma_{0}\left(  \frac{\partial}{\partial\tau}-\zeta\right)
-\sigma_{3}\left(  \nabla^{2}+\mu\right)  -\sigma_{1}\Delta_{\mathbf{x},\tau}.
\label{Nambu}%
\end{equation}
Here, $\sigma_{j}$ are the Pauli matrices. As far as the effective action
$S_{eff}$ is not a quadratic functional of the Hubbard-Stratonovich pair
field, the resulting functional integral over the pair field%
\begin{equation}
\mathcal{Z}\propto\int\mathcal{D}\Delta_{\mathbf{x},\tau}\mathcal{D}%
\bar{\Delta}_{\mathbf{x},\tau}\exp\left(  -S_{eff}\right)  \label{Z1}%
\end{equation}
cannot be calculated analytically exactly. As in the analogous problem in 3D
\cite{deMelo1993,TKD2008a,TKD2008b}, and as explained in the introduction, we
consider approximations provided by an expansion of the effective action
$S_{eff}$ over fluctuations of the pair field $\Delta_{\mathbf{x},\tau}$ about
its saddle-point value $\Delta$. The phase diagrams of a 2D Fermi gas in the
saddle-point approximation have been investigated in Refs.
\cite{Tempere2007,he}. The effective saddle-point action provides the
thermodynamic potential per unit area:%
\begin{equation}
\Omega_{sp}(T,\mu,\zeta;\Delta)=-\int\frac{d^{2}\mathbf{k}}{\left(
2\pi\right)  ^{2}}\left[  \frac{\ln\left(  2\cosh\beta E_{\mathbf{k}}%
+2\cosh\beta\zeta\right)  }{\beta}-\xi_{\mathbf{k}}\right]  -\frac{\Delta^{2}%
}{g}, \label{TP}%
\end{equation}
Here, $\xi_{\mathbf{k}}=k^{2}-\mu$ is the fermion energy, and $E_{\mathbf{k}%
}=\sqrt{\xi_{\mathbf{k}}^{2}+\Delta^{2}}$ is the Bogoliubov excitation energy.
The gap parameter $\Delta$ is determined from the gap equation generalized to
the imbalance case -- the minimum condition for the saddle-point thermodynamic
potential as a function of the gap parameter $\Delta$ at fixed temperature and
chemical potentials:%
\begin{equation}
\frac{\partial\Omega_{sp}\left(  \beta,\mu,\zeta;\Delta\right)  }%
{\partial\Delta}=0. \label{min}%
\end{equation}
For high temperatures ($T>T_{c}^{\ast}$) or at high levels of imbalance
($\zeta>\zeta_{c}$), thermodynamic potential will have its minimum at
$\Delta=0$, the unpaired normal state. Following the experimental observation
of superfluidity in imbalanced Fermi gases in 3D \cite{phasdiag3D}, the phase
diagram of the imbalanced Fermi gas has attracted a lot of attention (for a
recent review, see Ref. \cite{imbal3D}). To find the phase diagrams in 2D, the
above gap equation has to be solved in conjunction with the number equations
discussed in the remainder of this section.

\bigskip

\subsection{Gaussian fluctuations}

The next-order approximation brings into account fluctuations about the saddle
point:%
\[
\left\{
\begin{array}
[c]{c}%
\Delta_{\mathbf{x},\tau}=\Delta+\phi_{\mathbf{x},\tau},\\
\bar{\Delta}_{\mathbf{x},\tau}=\Delta+\bar{\phi}_{\mathbf{x},\tau}.
\end{array}
\right.
\]
We apply the Fourier expansion of the fluctuation coordinates:
\begin{align}
\phi_{\mathbf{x},\tau}  &  =\frac{1}{L\sqrt{\beta}}\sum_{\mathbf{q}}%
\sum_{n=-\infty}^{\infty}e^{i\mathbf{q\cdot r}-i\omega_{n}\tau}\varphi
_{\mathbf{k}}\left(  \omega_{n}\right)  ,\label{Fourier}\\
\bar{\phi}_{\mathbf{x},\tau}  &  =\frac{1}{L\sqrt{\beta}}\sum_{\mathbf{q}}%
\sum_{n=-\infty}^{\infty}e^{-i\mathbf{q\cdot r}+i\omega_{n}\tau}\bar{\varphi
}_{\mathbf{q}}\left(  \omega_{n}\right)  \label{Fourier1}%
\end{align}
where $L$ is the linear size of the 2D system, and $\omega_{n}=2\pi n/\beta$
(with $n=0,\pm1,\pm2,\ldots$) are the bosonic Matsubara frequencies. The
quadratic fluctuation contribution to the effective bosonic action is the
functional of complex fluctuation coordinates similar to that derived in Ref.
\cite{TKD2008b}:%

\begin{align}
S_{fl}  &  =\sum_{\mathbf{q}}\sum_{n=-\infty}^{\infty}\left\{  M_{1,1}\left(
q,i\omega_{n}\right)  \bar{\varphi}_{\mathbf{q}}\left(  \omega_{n}\right)
\varphi_{\mathbf{q}}\left(  \omega_{n}\right)  \right. \nonumber\\
&  +\frac{1}{2}M_{1,2}\left(  q,i\omega_{n}\right)  \left[  \bar{\varphi
}_{\mathbf{q}}\left(  \omega_{n}\right)  \bar{\varphi}_{-\mathbf{q}}\left(
\omega_{-n}\right)  \right. \nonumber\\
&  \left.  \left.  +\varphi_{\mathbf{q}}\left(  \omega_{n}\right)
\varphi_{-\mathbf{q}}\left(  \omega_{-n}\right)  \right]  \right\}  ,
\label{Sfl1}%
\end{align}
where $M_{j,k}\left(  q,i\omega_{n}\right)  $ are the matrix elements of the
inverse pair fluctuation propagator. They are determined by the expressions
(cf. Ref. \cite{TKD2008b}):%
\begin{align}
M_{1,1}\left(  \mathbf{q},i\omega_{n}\right)   &  =-\frac{1}{g}+\int
\frac{d^{2}\mathbf{k}}{\left(  2\pi\right)  ^{2}}\frac{X\left(  E_{\mathbf{k}%
}\right)  }{2E_{\mathbf{k}}}\nonumber\\
&  \times\left(  \frac{\left(  i\omega_{n}-E_{\mathbf{k}}+\xi_{\mathbf{k}%
+\mathbf{q}}\right)  \left(  E_{\mathbf{k}}+\xi_{\mathbf{k}}\right)  }{\left(
i\omega_{n}-E_{\mathbf{k}}+E_{\mathbf{k}+\mathbf{q}}\right)  \left(
i\omega_{n}-E_{\mathbf{k}}-E_{\mathbf{k}+\mathbf{q}}\right)  }\right.
\nonumber\\
&  \left.  -\frac{\left(  i\omega_{n}+E_{\mathbf{k}}+\xi_{\mathbf{k}%
+\mathbf{q}}\right)  \left(  E_{\mathbf{k}}-\xi_{\mathbf{k}}\right)  }{\left(
i\omega_{n}+E_{\mathbf{k}}-E_{\mathbf{k}+\mathbf{q}}\right)  \left(
i\omega_{n}+E_{\mathbf{k}+\mathbf{q}}+E_{\mathbf{k}}\right)  }\right)  ,
\label{M11}%
\end{align}
and%
\begin{align}
M_{1,2}\left(  \mathbf{q},i\omega_{n}\right)   &  =-\Delta^{2}\int\frac
{d^{2}\mathbf{k}}{\left(  2\pi\right)  ^{2}}\frac{X\left(  E_{\mathbf{k}%
}\right)  }{2E_{\mathbf{k}}}\nonumber\\
&  \times\left(  \frac{1}{\left(  i\omega_{n}-E_{\mathbf{k}}+E_{\mathbf{k}%
+\mathbf{q}}\right)  \left(  i\omega_{n}-E_{\mathbf{k}}-E_{\mathbf{k}%
+\mathbf{q}}\right)  }\right. \nonumber\\
&  \left.  +\frac{1}{\left(  i\omega_{n}+E_{\mathbf{k}}-E_{\mathbf{k}%
+\mathbf{q}}\right)  \left(  i\omega_{n}+E_{\mathbf{k}}+E_{\mathbf{k}%
+\mathbf{q}}\right)  }\right)  . \label{M22}%
\end{align}
Here, the following function has been introduced,%
\begin{equation}
X\left(  E_{\mathbf{k}}\right)  =\frac{\sinh(\beta E_{\mathbf{k}})}%
{\cosh(\beta E_{\mathbf{k}})+\cosh(\beta\zeta)}. \label{X}%
\end{equation}
The integration over fluctuation coordinates gives us the fluctuation
contribution $\Omega_{fl}(T,\mu,\zeta;\Delta)$ to the total grand-canonical
thermodynamic potential $\Omega$ per unit area:%
\begin{equation}
\Omega_{fl}(T,\mu,\zeta;\Delta)=\frac{1}{2\beta}\int\frac{d^{2}\mathbf{q}%
}{\left(  2\pi\right)  ^{2}}\sum_{n=-\infty}^{\infty}\ln\left[  \Gamma\left(
\mathbf{q},i\omega_{n}\right)  \right]  \label{fluct-tdpot}%
\end{equation}
with
\begin{align}
\Gamma\left(  \mathbf{q},i\Omega_{n}\right)   &  =M_{1,1}\left(
\mathbf{q},i\omega_{n}\right)  M_{1,1}\left(  \mathbf{q},-i\omega_{n}\right)
\nonumber\\
&  -M_{1,2}\left(  \mathbf{q},i\omega_{n}\right)  M_{1,2}\left(
\mathbf{q},-i\omega_{n}\right)  . \label{Gamma}%
\end{align}

\bigskip

\subsection{Number equations and the GPF approach \label{gpf}}

The fermion density and the density difference fix the chemical potentials
$\mu$ and $\zeta$ through the derivatives of the total thermodynamic potential
per unit area:%
\begin{align}
n  &  =-\left.  \frac{\partial\Omega}{\partial\mu}\right\vert _{T,\zeta
},\label{numes}\\
\delta n  &  =-\left.  \frac{\partial\Omega}{\partial\zeta}\right\vert
_{T,\mu}. \label{numes1}%
\end{align}
(remember that in our units $n=1/2\pi$). We can write out these equations by
splitting the total thermodynamic potential in saddle point and fluctuation
contributions.
\begin{align}
n  &  =-\left.  \frac{\partial\Omega_{sp}}{\partial\mu}\right\vert _{T,\zeta
}-\left.  \frac{\partial\Omega_{fl}}{\partial\mu}\right\vert _{T,\zeta
}\label{numeq-1}\\
\delta n  &  =-\left.  \frac{\partial\Omega_{sp}}{\partial\zeta}\right\vert
_{T,\mu}-\left.  \frac{\partial\Omega_{fl}}{\partial\zeta}\right\vert _{T,\mu}
\label{numeq-2}%
\end{align}
We will denote the first and second terms in the right hand side (RHS) of
expression (\ref{numeq-1}) for $n$ as $n_{sp}$ and $n_{fl}$, respectively.
Similarly, the terms in the RHS of expression (\ref{numeq-2}) will be denoted
by $\delta n_{sp}$ and $\delta n_{fl}$. Note that the thermodynamic potentials
obtained from expressions (\ref{TP}) and (\ref{fluct-tdpot}) are expressed as
a function not only of $T,\mu,\zeta$, but also of $\Delta$. This gap $\Delta$
is not an independent thermodynamic variable, and when considering
$\Omega(T,\mu,\zeta,\Delta)$ explicitly as a function of also $\Delta$, the
implicit dependence of $\Delta$ on the chemical potentials must be taken into
account in (\ref{numeq-1}),(\ref{numeq-2}):
\begin{align}
n  &  =-\left.  \frac{\partial\Omega_{sp}}{\partial\mu}\right\vert
_{T,\zeta,\Delta}-\left.  \frac{\partial\Omega_{fl}}{\partial\mu}\right\vert
_{T,\zeta,\Delta}-\left.  \frac{\partial\Omega_{fl}}{\partial\Delta
}\right\vert _{T,\zeta,\mu}\left.  \frac{\partial\Delta}{\partial\mu
}\right\vert _{T,\zeta},\nonumber\\
\delta n  &  =-\left.  \frac{\partial\Omega_{sp}}{\partial\zeta}\right\vert
_{T,\mu,\Delta}-\left.  \frac{\partial\Omega_{fl}}{\partial\zeta}\right\vert
_{T,\mu,\Delta}-\left.  \frac{\partial\Omega_{fl}}{\partial\Delta}\right\vert
_{T,\zeta,\mu}\left.  \frac{\partial\Delta}{\partial\zeta}\right\vert _{T,\mu
}. \label{numeq-GPF1}%
\end{align}
Note that the gap equation $\partial\Omega_{sp}/\partial\Delta=0$ at fixed
$T,\mu,\zeta$ implies that the implicit dependence of $\Delta$ on the chemical
potentials will only affect the fluctuation part of the thermodynamic
potential in the above equations. Different theories of the BEC-BCS crossover,
in any dimension, can be categorized by their choice of number and gap
equations. The simplest mean field approach only keeps the terms with
$\Omega_{sp}$. The Nozi\`{e}res and Schmitt-Rink approach also includes the
second terms in the RHS of expressions (\ref{numeq-GPF1}). Finally, the
Gaussian pair fluctuation approach includes also the last term in the RHS of
expressions (\ref{numeq-GPF1}). Note that in the literature, there is no
common opinion on which approach is best. For example, on the one hand,
Randeria \emph{et al}. \cite{R1,R2}, Hu \emph{et al}. \cite{Hu}, Keeling
\emph{et al}. \cite{Keeling} state that the derivative over $\mu$ must be
performed taking into account a variation of the gap determined by the gap
equation. On the other hand, Ohashi \emph{et al}.
\cite{Taylor2006,OhashiPRA67,Fukushima2007}, and Strinati \emph{et al}.
\cite{strinati,strinati2} use the other definition, considering $\Delta$ in
the number equations as an independent variable and, therefore, applying the
gap equation \emph{after} taking the derivatives $\partial\Omega/\partial\mu$.
In the papers \cite{Taylor2006,OhashiPRA67,Fukushima2007}, it is stated that
the last terms in (\ref{numeq-GPF1}) are the higher-order corrections with
respect to Gaussian quadratic fluctuations. Keeling \emph{et al}.
\cite{Keeling} correctly argue that both terms in those derivatives are of one
and the same order and emphasize that the existence of the second term is
crucial in two dimensions. Below, we demonstrate the key significance of
taking into account of the last terms (\ref{numeq-GPF1}) for the convergence
of fluctuation contributions to the fermion density in 2D.

\bigskip

As stated in the introduction, in order to treat the fluctuations, there is an
alternative to the Bogoliubov shift $\Delta_{\mathbf{x},\tau}=\Delta
+\phi_{\mathbf{x},\tau}$, namely the parametrization in amplitude and phase
fluctuations $\Delta_{\mathbf{x},\tau}\approx\Delta\left(  1+\delta
_{\mathbf{x},\tau}\right)  e^{i\theta_{\mathbf{x},\tau}}$. When, \emph{after}
this parametrization, the effective action is expanded with respect to
$\delta_{\mathbf{x},\tau}$ and $\theta_{\mathbf{x},\tau}$ (rather than
$\phi_{\mathbf{x},\tau}$ and $\bar{\phi}_{\mathbf{x},\tau}$), this leads to
another quadratic fluctuation action $S_{fl}^{\prime}$ which differs from
expression (\ref{Sfl1}) for $S_{fl}$ only by terms that vanish when applying
the gap equation. Correspondingly, the thermodynamic potentials $\Omega_{fl}$
and $\Omega_{fl}^{\prime}$ provided by those two actions lead to one and the
same contribution to the fermion density. Furthermore, keeping in
$S_{fl}^{\prime}$ only the leading order long-wavelength and low-energy terms
leads to the same effective \textquotedblleft hydrodynamic\textquotedblright%
\ action as in Ref. \cite{BKT-PRA2009}. In the particular case of a balanced
gas, the effective action of Ref. \cite{BKT-PRA2009} turns to the result of
Refs. \cite{Botelho2006,Babaev}. This means that the effective action
described as the result of the non-perturbative approach in Ref. \cite{Babaev}
can be equivalently re-derived within the perturbative NSR-like scheme.
Moreover, the present treatment can be considered as an extension of the
approach of Refs. \cite{Botelho2006,Babaev,BKT-PRA2009} beyond the
long-wavelength and low-energy approximation (and to imbalanced 2D gases). The
hydrodynamic action is particularly useful in extracting the superfluid
density $\rho_{s}$, by identifying it with the prefactor of the $(\nabla
\theta_{\mathbf{x},\tau})^{2}/2$ term in the expression for $S_{fl}^{\prime}$.
This identification yields straightforwardly\cite{BKT-PRA2009}:
\begin{equation}
\rho_{s}(T,\mu,\zeta,\Delta)=\frac{1}{4\pi}\int_{0}^{\infty}dk\text{ }k\left(
1-\frac{\xi_{k}}{E_{k}}X(E_{k})-k^{2}X^{\prime}(E_{k})\right)
\label{rhosupfl}%
\end{equation}
with $X(E_{k})$ given by expression (\ref{X}) and $X^{\prime}(E_{k})$ its
first derivative, evaluated in $E_{k}$. Once $\Delta,\mu,\zeta$ are obtained
for a given temperature (and interaction strength) by solving the gap and
number equations, they can be substituted in this expression to determine
whether the system is in the superfluid phase ($\rho_{s}\neq0$) or the normal
phase ($\rho_{s}=0$). As discussed in the results section, we also use this
expression to find the temperature $T_{BKT}$ of the phase transition between
those two states. Already we note that $\Delta=0$ leads to $\rho_{s}=0$, so
that $T_{BKT}<T_{c}^{\ast}$ and the superfluid state requires pair formation,
as it should.

\bigskip

\subsection{Pair fluctuation spectral functions}

From expression (\ref{TP}) for the saddle-point thermodynamic potential, we
derive the following expressions for the saddle-point densities:%
\begin{align}
n_{sp}  &  :=-\left.  \frac{\partial\Omega_{sp}}{\partial\mu}\right\vert
_{T,\zeta,\Delta}=\int\frac{d^{2}\mathbf{k}}{\left(  2\pi\right)  ^{2}}\left(
1-\frac{\xi_{\mathbf{k}}}{E_{\mathbf{k}}}\frac{\sinh\left(  \beta
E_{\mathbf{k}}\right)  }{\cosh\left(  \beta\zeta\right)  +\cosh\left(  \beta
E_{\mathbf{k}}\right)  }\right)  ,\label{deq1}\\
\delta n_{sp}  &  :=-\left.  \frac{\partial\Omega_{sp}}{\partial\zeta
}\right\vert _{T,\zeta}=\int\frac{d^{2}\mathbf{k}}{\left(  2\pi\right)  ^{2}%
}\frac{\sinh\left(  \beta\zeta\right)  }{\cosh\left(  \beta\zeta\right)
+\cosh\left(  \beta E_{\mathbf{k}}\right)  }. \label{deq2}%
\end{align}
Similarly, the fluctuation contributions to the fermion densities are
determined using (\ref{fluct-tdpot}) and using the GPF approach. The results
can be written as a sum over wavelengths and frequencies of pair fluctuation
structure factors:%
\begin{align}
n_{fl}  &  :=-\left.  \frac{\partial\Omega_{fl}}{\partial\mu}\right\vert
_{T,\zeta}=-\int\frac{d^{2}\mathbf{q}}{\left(  2\pi\right)  ^{2}}\frac
{1}{\beta}\sum_{n=-\infty}^{\infty}J\left(  \mathbf{q},i\omega_{n}\right)
,\label{n}\\
\delta n_{fl}  &  :=-\left.  \frac{\partial\Omega_{fl}}{\partial\zeta
}\right\vert _{T,\mu}=-\int\frac{d^{2}\mathbf{q}}{\left(  2\pi\right)  ^{2}%
}\frac{1}{\beta}\sum_{n=-\infty}^{\infty}K\left(  \mathbf{q},i\omega
_{n}\right)  . \label{dn}%
\end{align}
The pair fluctuation structure factors $J$ and $K$ are given by%
\begin{align}
J\left(  \mathbf{q},i\omega_{n}\right)   &  =\frac{1}{\Gamma\left(
\mathbf{q},i\omega_{n}\right)  }\left[  \frac{\partial M_{1,1}\left(
\mathbf{q},i\omega_{n}\right)  }{\partial\mu}M_{1,1}\left(  \mathbf{q}%
,-i\omega_{n}\right)  \right. \nonumber\\
&  \left.  -\frac{\partial M_{1,2}\left(  \mathbf{q},i\omega_{n}\right)
}{\partial\mu}M_{1,2}\left(  \mathbf{q},-i\omega_{n}\right)  \right]
,\label{J}\\
K\left(  \mathbf{q},i\omega_{n}\right)   &  =\frac{1}{\Gamma\left(
\mathbf{q},i\omega_{n}\right)  }\left[  \frac{\partial M_{1,1}\left(
\mathbf{q},i\omega_{n}\right)  }{\partial\zeta}M_{1,1}\left(  \mathbf{q}%
,-i\omega_{n}\right)  \right. \nonumber\\
&  \left.  -\frac{\partial M_{1,2}\left(  \mathbf{q},i\omega_{n}\right)
}{\partial\zeta}M_{1,2}\left(  \mathbf{q},-i\omega_{n}\right)  \right]  .
\label{K}%
\end{align}
We transform the Matsubara summations in (\ref{n}) and (\ref{dn}) to the
contour integrals in the complex plane as follows:%
\begin{equation}
\frac{1}{\beta}\sum_{n=-\infty}^{\infty}J\left(  \mathbf{q},i\omega
_{n}\right)  =\frac{1}{\pi}\int_{-\infty}^{\infty}\frac{\operatorname{Im}%
J\left(  \mathbf{q},\omega+i\delta\right)  }{e^{\beta\omega}-1}d\omega
,\;\delta\rightarrow+0. \label{Msum}%
\end{equation}
This allows to express the resulting fluctuation contributions to the fermion
density through the distribution functions for pair excitations:%
\begin{align}
n_{fl}  &  =\frac{1}{2\pi^{2}}\int_{0}^{\infty}g_{n}\left(  q\right)
qdq,\;\label{Nfls}\\
\delta n_{fl}  &  =\frac{1}{2\pi^{2}}\int_{0}^{\infty}g_{\delta n}\left(
q\right)  qdq.
\end{align}
The fluctuation distribution functions $g_{n}\left(  q\right)  $ and
$g_{\delta n}\left(  q\right)  $ are the integrals over the frequency with the
pair fluctuation structure factors:%
\begin{align}
g_{n}\left(  q\right)   &  =-\int_{-\infty}^{\infty}\frac{\operatorname{Im}%
J\left(  q,\omega+i\delta\right)  }{e^{\beta\omega}-1}d\omega,\label{gn}\\
g_{\delta n}\left(  q\right)   &  =-\int_{-\infty}^{\infty}\frac
{\operatorname{Im}K\left(  q,\omega+i\delta\right)  }{e^{\beta\omega}%
-1}d\omega. \label{gdn}%
\end{align}
The functions $g_{n}\left(  q\right)  $ and $g_{\delta n}\left(  q\right)  $
are proportional to the densities of states for the pair fluctuations. The
behavior of these functions is crucial for understanding of the pseudogap
properties and of different phase transitions in the imbalanced 2D Fermi gas.
In the NSR approach, the fluctuation distribution functions have a divergency,
and as a consequence no value of the chemical potential $\mu$ can be found so
that the number equation $n_{sp}+n_{fl}=n=1/(2\pi)$ is satisfied. In the GPF
approach, the divergency is overcome and the number equation can be satisfied.
In order to demonstrate this, we focus in the next section on the long
wavelength limit where exact analytic expressions for the distribution
functions are obtained.

\bigskip

\section{Distribution functions in the long-wavelength limit \label{expansion}%
}

\subsection{Long-wavelength expansion}

In order to investigate the problem of the long-wavelength convergence for the
fluctuation contributions to the density, it is necessary to derive
analytically the spectrum of low-lying and long-wavelength pair excitations.
For this purpose, we expand the matrix elements of the inverse pair
fluctuation propagator $M_{j,k}\left(  q,z\right)  $ in powers of $\left(
q,z\right)  $ up to the second-order terms in power of $z$ and $q$,%
\begin{align}
M_{1,1}\left(  \mathbf{q},z\right)   &  \approx A+Bq^{2}+Cz+Fz^{2}%
,\label{exp1}\\
M_{1,2}\left(  \mathbf{q},z\right)   &  \approx D+Eq^{2}+Hz^{2}. \label{exp2}%
\end{align}
The derivation of the coefficients is rather tedious. In this connection, here
only the final results are represented. The coefficients in the $M_{1,1}$
expansion are given by%
\begin{equation}
A=\frac{1}{2\pi}\int_{0}^{\infty}kdk\left(  \frac{1}{2k^{2}-E_{b}}-\frac
{E_{k}^{2}+\xi_{k}^{2}}{4E_{k}^{3}}X\left(  E_{\mathbf{k}}\right)
-\frac{\Delta^{2}}{4}\frac{X^{\prime}\left(  E_{k}\right)  }{E_{k}^{2}%
}\right)  ,
\end{equation}%
\begin{align}
B  &  =\frac{1}{16\pi}\int_{0}^{\infty}kdk\frac{k^{2}\left(  E_{k}^{4}%
+7E_{k}^{2}\xi_{k}^{2}-10\xi_{k}^{4}\right)  -E_{k}^{4}\xi_{k}+3\xi_{k}%
^{3}E_{k}^{2}}{E_{k}^{7}}X\left(  E_{\mathbf{k}}\right) \nonumber\\
&  +\frac{\Delta^{2}}{8\pi}\int_{0}^{\infty}kdk\left(  \frac{\xi_{k}\left(
E_{k}^{2}-3k^{2}\xi_{k}\right)  }{E_{k}^{6}}X^{\prime}\left(  E_{k}\right)
\right. \nonumber\\
&  \left.  +\frac{\xi_{k}\left(  3k^{2}\xi_{k}-E_{k}^{2}\right)  -k^{2}%
E_{k}^{2}}{2E_{k}^{5}}X^{\prime\prime}\left(  E_{k}\right)  -\frac{k^{2}%
\xi_{k}^{2}}{3E_{k}^{4}}X^{(3)}\left(  E_{k}\right)  \right)  ,
\end{align}%
\begin{equation}
C=-\frac{1}{8\pi}\int_{0}^{\infty}kdk\frac{\xi_{k}}{E_{k}^{3}}X\left(
E_{\mathbf{k}}\right)  -\frac{\Delta^{2}}{8\pi}\int_{0}^{\infty}%
kdk\frac{X^{\prime}\left(  E_{k}\right)  }{\xi_{k}E_{k}^{2}},
\end{equation}%
\begin{equation}
F=-\frac{1}{32\pi}\int kdk\frac{E_{k}^{2}+\xi_{k}^{2}}{E_{k}^{5}}X\left(
E_{\mathbf{k}}\right)  .
\end{equation}
and those in the $M_{1,2}$ expansion are given by%
\begin{equation}
D=\frac{\Delta^{2}}{8\pi}\int_{0}^{\infty}kdk\frac{X\left(  E_{\mathbf{k}%
}\right)  -E_{\mathbf{k}}X^{\prime}\left(  E_{\mathbf{k}}\right)
}{E_{\mathbf{k}}^{3}},
\end{equation}%
\begin{align}
E  &  =\frac{\Delta^{2}}{16\pi}\int_{0}^{\infty}kdk\frac{10k^{2}\xi_{k}%
^{2}-3E_{k}^{2}\left(  \xi_{k}+k^{2}\right)  }{E_{\mathbf{k}}^{7}}X\left(
E_{\mathbf{k}}\right) \nonumber\\
&  +\frac{\Delta^{2}}{8\pi}\int_{0}^{\infty}kdk\left(  \frac{\xi_{k}E_{k}%
^{2}+k^{2}E_{k}^{2}-3k^{2}\xi_{k}^{2}}{2E_{k}^{6}}\left[  2X^{\prime}\left(
E_{k}\right)  -E_{k}X^{\prime\prime}\left(  E_{k}\right)  \right]  \right.
\nonumber\\
&  \left.  -\frac{\xi_{k}^{2}k^{2}}{3E_{k}^{4}}X^{(3)}\left(  E_{k}\right)
\right)  ,
\end{align}%
\[
H=\frac{\Delta^{2}}{32\pi}\int_{0}^{\infty}kdk\frac{X\left(  E_{\mathbf{k}%
}\right)  }{E_{k}^{5}}.
\]
In these expressions $X^{\prime}$,$X^{\prime\prime}$ and $X^{(3)}$ are the
first, second and third derivatives of the function $X$ given by expression
(\ref{X}), with respect to its argument. In the literature, an analogous
expansion was performed for 3D at finite temperatures in Refs.
\cite{Taylor2006,OhashiPRA67,Fukushima2007} in the strong-coupling limit, and
in Ref. \cite{R1} at low temperatures. The present expansion is all-coupling
and all-temperature, because no restriction is imposed on the thermodynamic parameters.

\subsection{Structure factor}

Let us substitute the expansions (\ref{exp1}), (\ref{exp2}) to the structure
factor $J\left(  \mathbf{q},z\right)  $ in order to obtain its long-wavelength
and low-energy form $J_{lw}\left(  \mathbf{q},z\right)  $. The spectrum of
pair bosonic excitations is determined by the poles of $J_{lw}\left(
\mathbf{q},z\right)  $. In the long-wavelength and low-energy range, these
roots are $z=\pm\omega_{q}$ with the pair excitation frequency $\omega_{q}$
which satisfies the Goldstone theorem,%
\begin{equation}
\omega_{q}=q\sqrt{v^{2}+\kappa^{2}q^{2}} \label{freq1}%
\end{equation}
with the parameters (cf. the analogous expansion in the 3D case, Ref.
\cite{KTD-JLTP2011})%
\begin{align}
v  &  =\sqrt{\frac{2A\left(  B-E\right)  }{C^{2}+2A\left(  H-F\right)  }%
},\label{pars}\\
\kappa &  =\sqrt{\frac{C^{2}\left(  B-E\right)  \left(  4A\left(
BH-EF\right)  +C^{2}\left(  B+E\right)  \right)  }{\left(  C^{2}+2A\left(
H-F\right)  \right)  ^{3}}}. \label{pars2}%
\end{align}
The parameter $v$ has the dimensionality of velocity and tends to the first
sound velocity in the low-temperature limit . As far as the parameter $D$ is
proportional to $\Delta^{2}$, the velocity parameter for pair excitations
turns to zero at the phase boundary when $\Delta=0$. In the BEC limit, when
$E_{b}\gg1$, we find that $\mu\rightarrow-E_{b}/2$ and $\kappa\rightarrow1/4$.
This results in the pair excitation spectrum $\omega_{q}\rightarrow q^{2}/2$
at the BEC side.

In order to calculate the long-wavelength distribution function, we keep the
lowest-order terms in powers of $z$ and $q^{2}$ in the numerator of
$J_{lw}\left(  \mathbf{q},z\right)  $. This gives us the result%
\begin{equation}
J_{lw}\left(  \mathbf{q},z\right)  =\frac{a_{q}}{z-\omega_{q}}+\frac{b_{q}%
}{z+\omega_{q}}, \label{jlw1}%
\end{equation}
where the coefficients $a_{q}$ and $b_{q}$ are related to the constants
determined above as%
\begin{equation}
a_{q}=\frac{\alpha+\lambda\omega_{q}+\chi q^{2}}{2\omega_{q}\left[
C^{2}+2A\left(  H-F\right)  \right]  },\;b_{q}=-\frac{\alpha-\lambda\omega
_{q}+\chi q^{2}}{2\omega_{q}\left[  C^{2}+2A\left(  H-F\right)  \right]  }
\label{coefs}%
\end{equation}
with the notations%
\begin{equation}
\alpha=D_{\mu}D-A_{\mu}A,\; \lambda=A_{\mu}C-C_{\mu}A,\; \chi=E_{\mu}D+D_{\mu
}E-A_{\mu}B-B_{\mu}A. \label{cfs1}%
\end{equation}
Here $A_{\mu},B_{\mu},\ldots$ are the derivatives $A_{\mu}\equiv\partial
A/\partial\mu$, etc. The distribution function is calculated setting
$z=\omega+i\delta$ with $\delta\rightarrow+0$. This gives us the structure
factor as a superposition of the delta functions. The distribution function
then takes the form%
\begin{equation}
g_{n}^{\left(  lw\right)  }\left(  q\right)  =\frac{1}{4\pi}\frac{\lambda
}{C^{2}+2A\left(  H-F\right)  }\left[  \frac{\alpha+\chi q^{2}}{\lambda
\omega_{q}}\coth\left(  \frac{\beta\omega_{q}}{2}\right)  -1\right]  .
\label{gn1}%
\end{equation}
For the other distribution function, $g_{\delta n}^{\left(  lw\right)
}\left(  q\right)  $, the derivations are the same, but with a replacement of
the derivatives over $\mu$ by the corresponding derivatives over $\zeta$.

\subsection{Distribution functions in the paired state}

Here we consider the paired state of the quasicondensate in which the gap
parameter $\Delta\neq0$. In this case, the gap parameter obeys the gap
equation:%
\begin{equation}
\frac{1}{4\pi}\int_{0}^{\infty}\frac{X\left(  E_{\mathbf{k}}\right)
}{E_{\mathbf{k}}}kdk+\frac{1}{g}=0. \label{gapeq}%
\end{equation}
The strength $g$ of the contact interaction is expressed through the
two-particle binding energy $E_{b}$ in 2D by the equation (\ref{renorm}). The
difference of coefficients $A-D$ is proportional to the left hand side (LHS)
of the gap equation. Therefore, as long as the gap equation is satisfied, we
obtain $D=A$. Moreover, because the derivatives of matrix elements are
calculated while keeping the gap equation satisfied, we find that $A_{\mu
}=D_{\mu}\;$and$\;A_{\zeta}=D_{\zeta}.$This implies in particular that the
coefficient $\alpha=0$, as is evident from expression (\ref{cfs1}), and we
find:%
\begin{equation}
g_{n}^{\left(  lw\right)  }\left(  q\right)  =\frac{1}{4\pi}\frac{\lambda
}{C^{2}+2A\left(  H-F\right)  }\left[  \frac{\chi}{\lambda}\frac{q^{2}}%
{\omega_{q}}\coth\left(  \frac{\beta\omega_{q}}{2}\right)  -1\right]  .
\label{gn2}%
\end{equation}
Thus $g_{n}^{(lw)}\left(  q\right)  $ tends to a finite value at
$q\rightarrow0$ for $\Delta\neq0,$ and behaves as $q^{-2}$ at $q\rightarrow0$
for $\Delta=0$. As a result, the fluctuation contributions $n_{fl}$ and
$\delta n_{fl}$ in 2D are \emph{finite} at $\Delta\neq0$. They can diverge
only at $\Delta=0$. This is to be contrasted with the NSR scheme, where the
order parameter $\Delta$ is treated as independent variable, and where
$n_{fl}$ and $\delta n_{fl}$ in 2D diverge \emph{for all} $\Delta$.
\begin{figure}
[h]
\begin{center}
\includegraphics[
height=11.3732cm,
width=8.6436cm
]%
{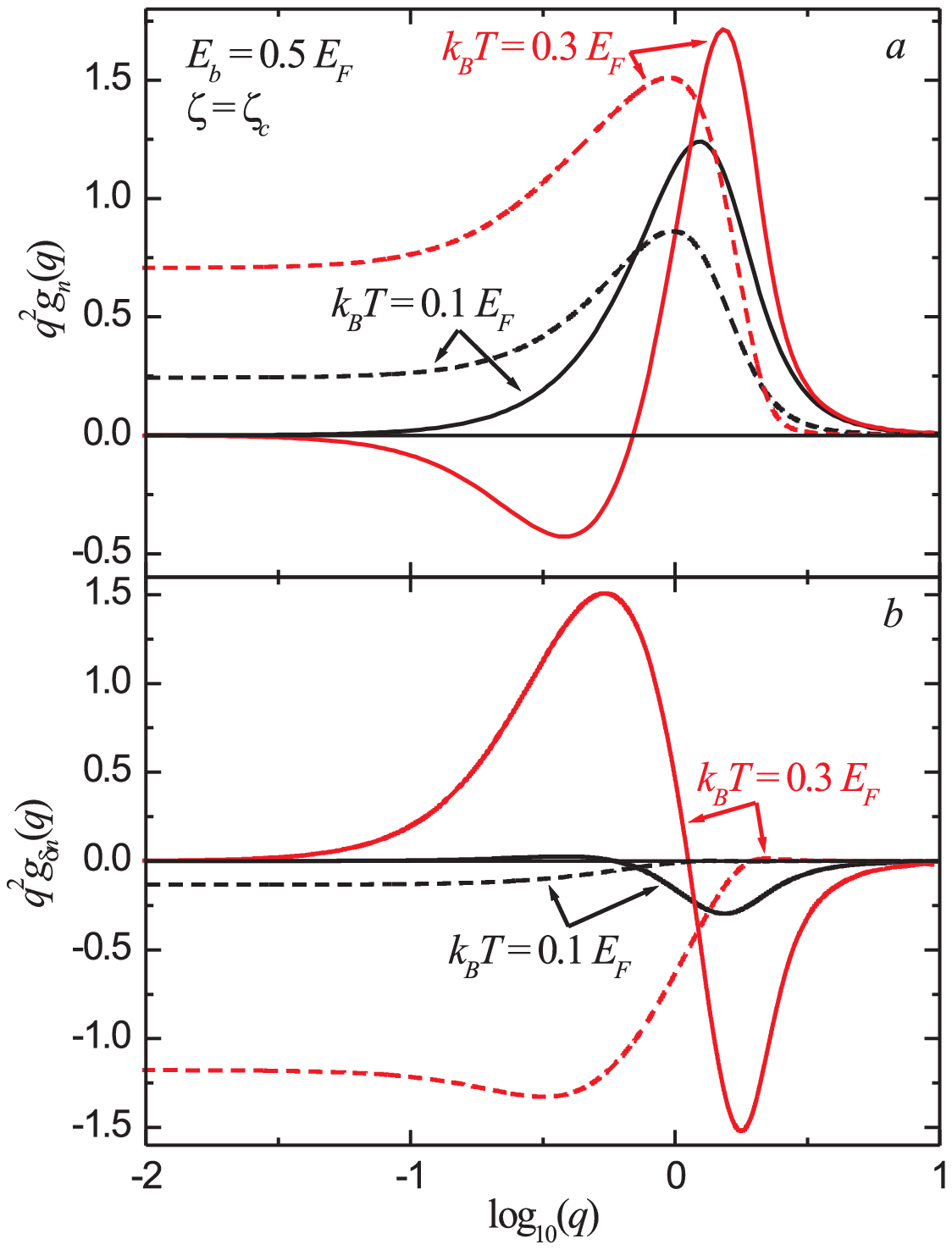}%
\caption{(Color online) Distribution functions for (\emph{a}) the fluctuation
contribution to the fermion density and (\emph{b}) the density difference, at
the binding energy $E_{b}=0.5E_{F}$. The solid and dashed curves show the
spectral functions obtained, respectively, within the GPF formalism and within
the standard NSR scheme. The spectral functions are calculated for critical
values $\zeta=\zeta_{c}$ of the chemical potential imbalance, and for two
different temperatures. In the graphs, the spectral functions are multiplied
by $q^{2}$ in order to show clearly their behavior at small $q$.}%
\end{center}
\end{figure}

Fig. 1 shows the behavior of the fluctuation distribution functions
$g_{n}\left(  q\right)  $ and $g_{\delta n}\left(  q\right)  $ for different
temperatures, at binding energy $E_{b}=0.5$ and at the critical value of the
chemical potential imbalance $\zeta=\zeta_{c}\left(  E_{b},T\right)  $. The
critical value $\zeta_{c}$ for a given $\left(  E_{b},T\right)  $ is
determined as the highest imbalance at which the order parameter $\Delta$ is
other than zero. The dashed lines correspond to the NSR scheme and reveal a
$q^{-2}$ long wavelength divergence. The full lines show the results in the
GPF scheme, where the long-wavelength divergence is absent. This behavior is
seen both for $g_{n}(q)$ and $g_{\delta n}(q)$. In the limit $\Delta
\rightarrow0$, the functions $qg_{n}\left(  q\right)  $ and $qg_{\delta
n}\left(  q\right)  $ become logarithmically divergent. However, the sign of
this divergence is \emph{opposite} to that of the divergence of the functions
calculated neglecting the variation of $\Delta$.%

\begin{figure}
[h]
\begin{center}
\includegraphics[
height=13.1376cm,
width=10.5344cm
]%
{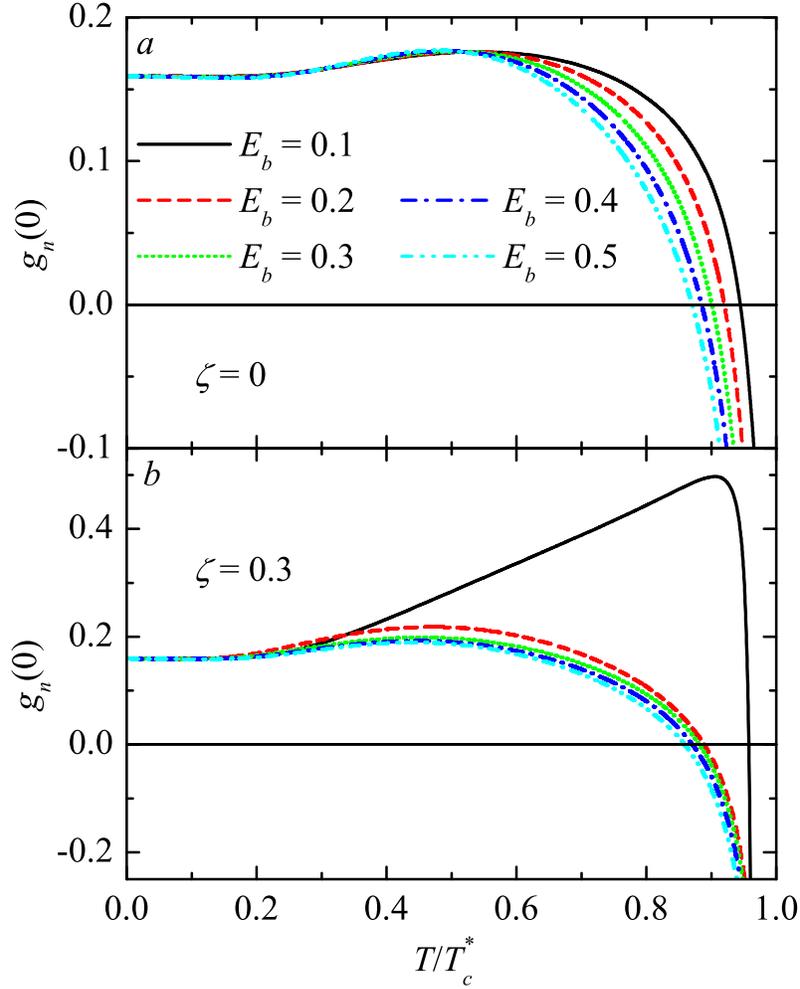}%
\caption{(Color online) The fluctuation distribution function $g_{n}\left(
0\right)  $ as a function of temperature for $\zeta=0$ (\emph{a}) and for
$\zeta=0.3$ (\emph{b}). The values of the binding energy are shown in the
figure.}%
\end{center}
\end{figure}

For the lower temperature shown in Fig. 1, $T/T_{F}=0.1$, $g_{n}(q)$ remains
positive, whereas for the higher temperature $T/T_{F}=0.3$, there is a sign
change in $g_{n}(q)$ as it becomes negative for small $q$. Regions of negative
value for the fluctuation distribution function $g_{\delta n}(q)$ are
expected, as the sign will change depending on which species is the majority
species. However $g_{n}(q)$ is expected to remain positive, as it is
proportional to the pair fluctuation density of states. The appearance of a
long-wavelength instability heralds the breakdown of the paired state. We can
track the onset of this instability by studying $g_{n}(q\rightarrow0)$ as a
function of temperature. At low temperatures, $g_{n}(0)$ is positive, and the
fluctuation density function remains positive for all $q$. At high
temperatures $g_{n}(0)$ becomes negative, signalling the long-wavelength
instability. We denote the temperature separating the two regions by $T_{p}$,
and find this temperature through solving $g_{n}(0)=0$ with respect to
temperature. The behavior of $g_{n}(0)$ as a function of temperature is shown
in Fig.2, for different values of the binding energy $E_{b}$ and both for
balanced and imbalanced systems. The function $g_{n}\left(  0\right)  $
diverges when the temperature achieves the limit $T=T_{c}^{\ast}$ at which
$\Delta=0$. This result explicitly follows from the analytic properties of the
long-wavelength expansion of the distribution functions as discussed above.
The temperature $T_{p}$ at which $g_{n}(0)=0$ lies below the temperature
$T_{c}^{\ast}$ where we find $\Delta=0$, and above the critical temperature
$T_{BKT}$ for superfluidity.

The temperature $T_{p}$ does not correspond to a phase transition, because the
gap equation is satisfied with a finite density both below and above $T_{p}$.
Nevertheless, because $T_{p}$ is the temperature at which the fluctuation
density of states changes it qualitative behavior, we hypothesize that $T_{p}$
corresponds to a crossover between the normal and pseudogap states. This will
be further substantiated by comparing our spectral functions to the
experimental ones in sec. \ref{Experiment}. The joint solution of the gap and
number equations within the GPF theory then formally provides a non-superfluid
quasicondensate at temperatures below $T_{p}$. Indeed, for temperatures
$T_{BKT}<T<T_{p}$ the phase coherence is destroyed by the phase fluctuations
according to the BKT mechanism, resulting in the phase fluctuating
quasicondensate discussed by Kagan \cite{Kagan}. Through the interpretation of
the spectral function, we will denote this temperature region as the
\textquotedblleft pseudogap regime\textquotedblright. It is worth noting that
the total fermion density within the GPF theory is finite at $T=T_{p}$ without
the necessity to introduce any cutoff in the integrals over $q$. In the next
section we set up phase diagrams identifying the regions where the superfluid
phase and the non-coherent paired phase occur.

\bigskip

\section{Phase diagrams \label{phasediagrams}}

In order to get the complete set of equations for phase diagrams, the number
equations (\ref{numeq-GPF1}),(\ref{numeq-GPF1}) and the generalized gap
equation (\ref{min}) are solved jointly with the equation for the BKT
transition temperature $T_{BKT}$ determined by \cite{Nelson1977}%
\begin{equation}
T_{BKT}-\frac{\pi}{2}\rho_{s}\left(  T_{BKT}\right)  =0, \label{NK}%
\end{equation}
where $\rho_{s}$ is the superfluid pair density given by Eq. (\ref{rhosupfl}).
To investigate the phase transitions for the Fermi gas in 2D for different
binding energies, we have calculated the critical temperatures of the BKT
phase transition $T_{BKT}$ and the critical temperature $T_{p}$ below which
the phase fluctuating quasicondensate is formed, as a function of the binding
energy $E_{b}$. Because the fluctuation contribution to the density is finite
at $T_{p}$ and at $T_{BKT}$ , these temperatures can be self-consistently
determined from the joint solution of the gap and number equations with the
\emph{complete} thermodynamic potential $\Omega=\Omega_{sp}+\Omega_{fluct}$.%

\begin{figure}
[h]
\begin{center}
\includegraphics[
height=4.6289in,
width=3.2591in
]%
{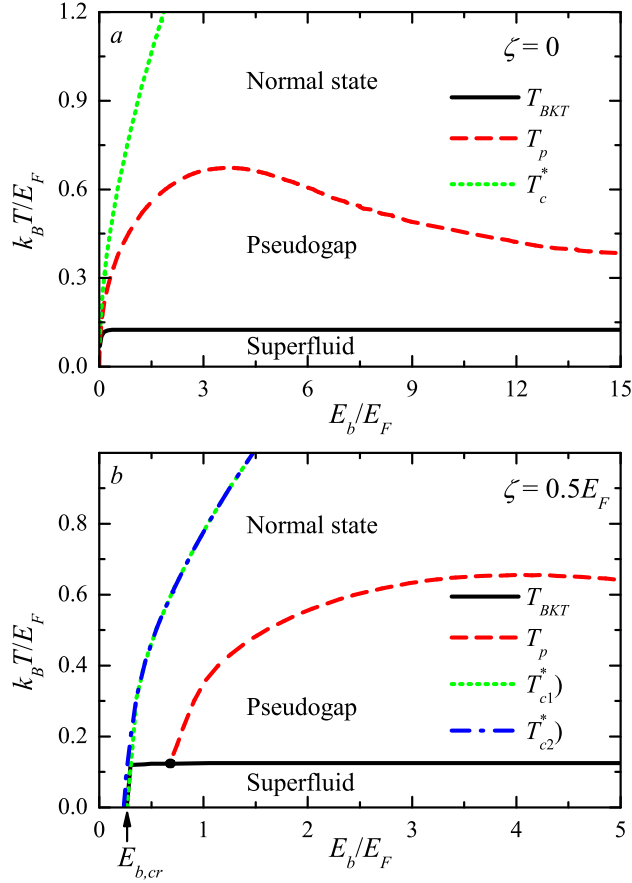}%
\caption{(Color online) Phase diagrams for the Fermi gas in 2D (\emph{a}) in
the case of equal spin up and spin down populations, (\emph{b}) for the
imbalanced Fermi gas with the chemical potential imbalance $\zeta=0.5$. The
crossover pairing temperature $T_{p}$ for the pseudogap formation and the BKT
transition temperature $T_{BKT}$ are shown with solid and dot-dashed and solid
curves, respectively. The dashed and dotted curves show the mean-field phase
transition temperatures $T_{c1}^{\ast},T_{c2}^{\ast}$ (explained in the text).
The arrow indicates the lowest binding energy at nonzero imbalance when
preformed pairs can arise.}%
\end{center}
\end{figure}

The phase diagrams in Fig. 3 show the critical temperatures for cold fermions
in 2D as a function of the binding energy $E_{b}$ for the balanced case (panel
$a$) and for the chemical potential imbalance $\zeta=0.5$ (panel $b$). The
formation of the superfluid state is indicated by the critical temperature
$T_{BKT}$ of the BKT phase transition. The pseudogap temperature $T_{p}$ is
the upper bound for the existence of the phase fluctuating quasicondensate
described in the previous section. We also show the mean-field temperature for
pair formation, $T_{c}^{\ast}$, obtained by solving gap and number equations
with $\Omega=\Omega_{sp}$. According to Ref. \cite{RanderiaNat} (for 3D), the
unitary gas can exist in the normal state with pairing correlations called
preformed pairs which survive at temperatures up to this $T_{c}^{\ast}$. The
critical temperatures for the balanced case, Fig. 3~(\emph{a}), were
calculated in Ref. \cite{QFS2010}. Here, they are reproduced in order to
compare them with those for a nonzero imbalance. The population imbalance
brings new features to the phase diagram: a phase separation region (between
$T_{c1}^{\ast}$ and $T_{c2}^{\ast}$) and a minimum binding energy $E_{b,cr}$
required for superfluidity.

For the balanced Fermi gas the superfluid state exists for any value of the
binding energy: the BKT critical temperature as well as other critical
temperatures gradually decrease with decreasing $E_{b}$, remaining always
finite. However, when $\zeta\neq0$ a minimum value of the binding energy
$E_{b,cr}$ is required for superfluidity to exist even at $T=0$. As shown in
Fig. 3, the pseudogap temperature $T_{p}$ does not grow unboundedly when
increasing the binding energy $E_{b}$. For $\zeta=0.5$ it achieves its maximum
at around $E_{b}\approx4$ and then slowly decreases tending to a finite value.
Consequently, in the strong-coupling regime the pseudogap temperature is
suppressed with respect to the mean-field prediction, where it is often
identified with our pair formation temperature $T_{c}^{\ast}$, as in
\cite{RanderiaNat}. This behavior is qualitatively similar to that for the
critical temperature $T_{c}$ as a function of $1/a_{s}$ for the cold fermions
in 3D obtained first in Ref. \cite{deMelo1993} accounting for the Gaussian fluctuations.

The critical temperatures $T_{c1}^{\ast}$ and $T_{c2}^{\ast}$ coincide with
each other in the balanced case, and they can be different in the imbalanced
case: the area between $T_{c1}^{\ast}$ and $T_{c2}^{\ast}$ is the
\textquotedblleft phase-separated state\textquotedblright. In the
phase-separated state, uniform phases are not possible. The temperatures
$T_{c1}^{\ast}$ and $T_{c2}^{\ast}$ were already calculated in Ref.
\cite{BKT-PRA2009}. The temperature $T_{p}$ is determined for the state with
$\Delta\neq0$. Therefore a non-zero imbalance does not lead to a splitting of
this critical temperature. However, a tricritical point appears at
$T_{p}=T_{BKT}$ in the phase diagram joining three regions: the superfluid
state, the pseudogap regime and the normal state. This tricritical point is
rather conventional as far as the pseudogap temperature indicates a crossover
rather than a sharp transition.%

\begin{figure}
[h]
\begin{center}
\includegraphics[
height=2.7905in,
width=3.6843in
]%
{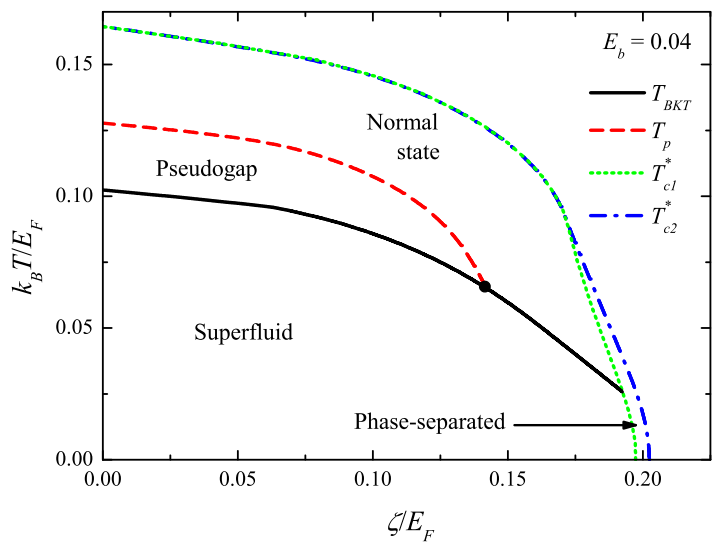}%
\caption{(Color online) Finite-temperature phase diagram for cold fermions in
2D in the variables $\left(  T,\zeta\right)  $ for the binding energy
$E_{b}=0.04E_{F}$. The full dot indicates a tricritical point.}%
\end{center}
\end{figure}

At zero imbalance, $T_{p}>T_{BKT}$, and the phase coherence in the range
$T_{BKT}<T<T_{p}$ is destroyed by phase fluctuations that lead to a phase
fluctuating quasicondensate. However, at nonzero imbalance, there is a region
where pseudogap temperature crosses the BKT temperature for superfluidity.
This result is interesting in connection with recent experiments on
high-$T_{c}$ superconductors \cite{Rourke}, that show a crossing of the
zero-field superconducting transition temperature and the temperature
indicating the opening of the pseudogap in overdoped La$_{2-x}$Sr$_{x}%
$CuO$_{4}$. The crossing of pseudogap temperature and BKT temperature is also
seen in Fig. 4, showing the phase diagram in the variables $\left(
T,\zeta\right)  $, for the binding energy $E_{b}/E_{F}=0.04$. Here, the same
critical temperatures and phase regions are identified as in Fig. 3(\emph{b}).
Increasing imbalance is not only detrimental to the superfluid phase, it also
suppresses the pseudogap regime.

\bigskip

\section{Comparison with experiment \label{Experiment}}

In the experiment \cite{Feld} on pairing of cold fermions in two dimensions,
the single-particle spectral function $A\left(  \mathbf{q},\omega\right)  $ is
measured for different values of the wave number $q$. The spectral function
exhibits peaks whose positions indicate the energies of the pair excitations.
In the strong-coupling regime, these energies are close to the pair binding
energy $E_{b}$. However, as stated in the paper, some discrepancies remain
between the peak positions observed in the experiment and those predicted by
the mean-field theory. The deviation \textquotedblleft could stem from beyond
mean-field effects provoked by our two-dimensional geometry and interaction
energy shifts\textquotedblright\ \cite{Feld}.

In the GPF approach, the pair fluctuation contribution of the fermion density
is expressed through the integral (\ref{gn}), where the structure factor
$J\left(  \mathbf{q},\omega\right)  $ describes the spectrum of the pair
excitations of the fermion system. Thus there should be a correspondence of
the peaks of the structure factor $J\left(  \mathbf{q},\omega\right)  $ with
the peaks of the spectral function $A\left(  \mathbf{q},\omega\right)  $. In
this connection, we compare the positions of the peaks of the spectral
function measured in Ref. \cite{Feld} with those of the structure factor
calculated within the GPF approach. The results are shown in Fig. 5 for $q=0$
and $T/T_{F}=0.27$, where $k_{B}T_{F}\equiv E_{F}$. The 2D scattering length
$a_{2D}$ is related to the binding energy $E_{b}$ as $a_{2D}=\hbar
/\sqrt{mE_{b}}$. The value of the Fermi wave vector taken from Ref.
\cite{Feld} is $k_{F}=8.1%
\operatorname{\mu m}%
^{-1}$. When using the mass of the fermion atom $m\approx39.964%
\operatorname{u}%
$, we found that the frequency $\nu_{F}\equiv E_{F}/\left(  2\pi\hbar\right)
$ corresponding to the Fermi energy is $\nu_{F}=8.2967%
\operatorname{kHz}%
$.

For the visualization of the peaks of the structure factor, we have used
$J\left(  \mathbf{q},\omega+i\gamma\right)  $ with a finite damping parameter
$\gamma$ (as in Refs. \cite{TKD2008a,TKD2008b}, where this parameter was
introduced to facilitate the numeric calculations). Here, the value
$\gamma=0.2\pi/\beta$ is used, where $\beta=1/\left(  k_{B}T\right)  $ is the
inverse temperature.

The parameters of the state (the chemical potential $\mu$ and the gap
parameter $\Delta$) are determined for each plot from the joint solution of
the gap and number equations. In the number equation, the Gaussian
fluctuations are included within the GPF formalism. The GPF method provides a
finite (convergent) pair fluctuation contribution for any finite $\Delta$
without any cutoff for the pair momentum. This is to be contrasted with the
standard NSR scheme which leads to a divergence of the fluctuation
contribution at any $\Delta$. Therefore the standard NSR scheme cannot be used
for the description of the pseudogap state, whereas the GPF approach can
describe this regime.%

\begin{figure}
[h]
\begin{center}
\includegraphics[
height=2.9024in,
width=5.4844in
]%
{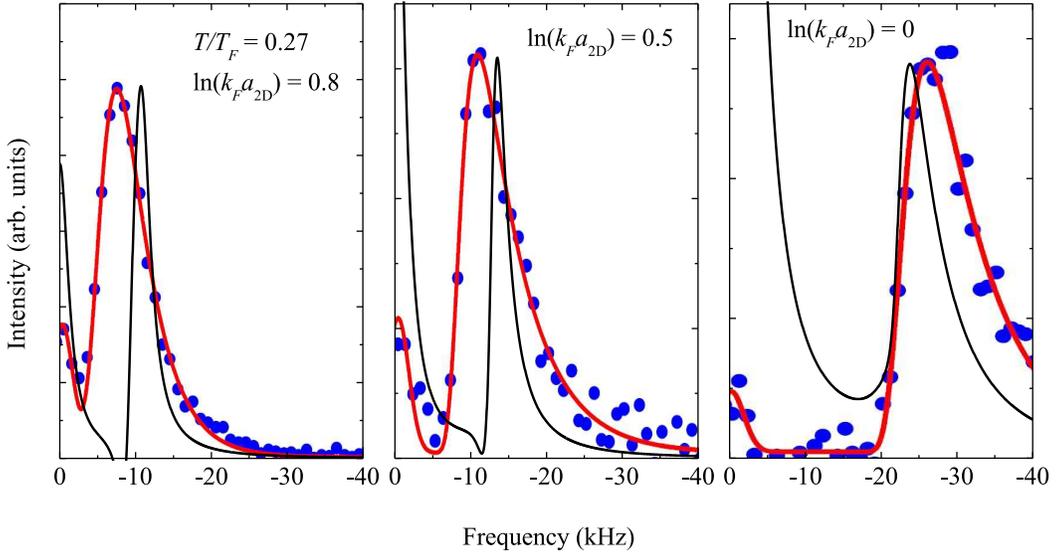}%
\caption{\emph{Full dots}: measured energy distribution curve $A\left(
q=0,\omega\right)  $ for $\ln(k_{F}a_{2D})=0.8$ from Ref. \cite{Feld}.
\emph{The red solid line} is the fit by elementary functions to the
experimental data performed in Ref. \cite{Feld}. \emph{The black solid line}:
the structure factor $J\left(  q=0,\omega\right)  $ calculated in the present
work within the GPF approach.}%
\end{center}
\end{figure}

In Fig. 5, the high peak at $\omega=0$ in our results has no relation to the
energies of the pair excitations: it is an intrinsic feature of the structure
factor. The other peak of our structure factor at $\omega<0$ is positioned
remarkably close to the measured peak of the spectral function attributed to
the pair excitation energy in Ref. \cite{Feld}, especially for the relative
high coupling strength at $\ln\left(  k_{F}a_{2D}\right)  =0$. A possible
reason for the remaining difference between the peak positions of the
calculated structure factor $J\left(  \mathbf{q},\omega\right)  $ and the
measured peak spectra can be the experimental uncertainty in the determination
of the Fermi wave vector, which can be slightly different from the reported
value $k_{F}=8.1%
\operatorname{\mu m}%
^{-1}$. Another possible source of the remaining difference is the similar
experimental uncertainty on $\ln(k_{F}a_{2D})$. In particular, this
uncertainty can be provided by the facts that the Fermi wave vector determined
in Ref. \cite{Feld} is a trap-averaged\ rather than local quantity. It should
be noted that the structure factor $J\left(  \mathbf{q},\omega\right)  $
calculated with the mean-field values for $\mu$ and $\Delta$ leads to a large
discrepancy between the peaks of $J\left(  \mathbf{q},\omega\right)  $ and
those of the measured spectral function. This confirms the importance of
including fluctuations through the GPF approach for the description of the
pseudogap state of cold fermions in 2D.%

\begin{figure}
[h]
\begin{center}
\includegraphics[
height=3.6991in,
width=2.4058in
]%
{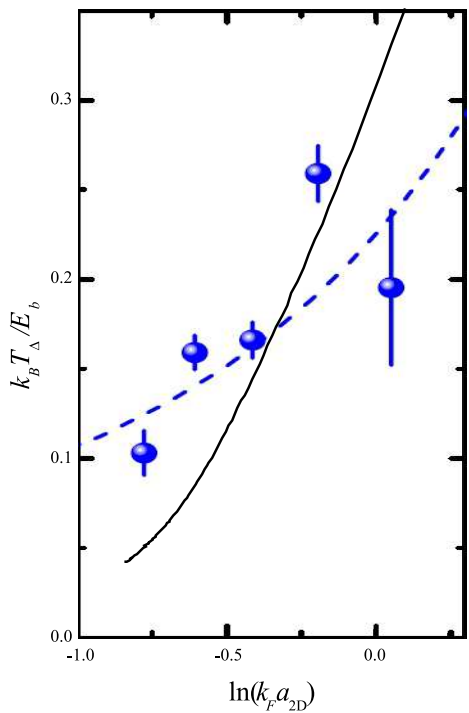}%
\caption{\emph{Solid curve}: calculated pseudogap pairing temperature $T_{p}$
(in units of $E_{b}/k_{B}$) compared with the experimentally \cite{Feld}
determined pairing temperature $T_{B}^{\ast}$ (full dots). \emph{Dotted
curve}: the mean-field critical temperature $T_{c}^{\ast}$ scaled by the
factor 0.36.}%
\end{center}
\end{figure}

In Ref. \cite{Feld}, the pairing crossover temperature $T^{\ast}$\ and the
pseudogap pairing temperature $T_{B}^{\ast}<T^{\ast}$ have been introduced.
The temperature $T^{\ast}$ coincides with the mean-field transition
temperature $T_{c}^{\ast}$. The temperature $T_{B}^{\ast},$ as stated in Ref.
\cite{Feld}, indicates the formation of pairs, and has the same physical
meaning as the temperature $T_{p}$ obtained in our study. As far as the
transition between the normal and paired states is a crossover rather than a
true phase transition, the pairing temperatures $T_{B}^{\ast}$ and $T_{p}$
only approximately indicate the formation of a paired state. In Fig. 6, the
pseudogap pairing temperature $T_{p}$ is compared with the experimental data
for $T_{B}^{\ast}$. The dotted curve shows the scaled mean-field transition
temperature from Ref. \cite{Feld}; the scaling indicates that the experimental
result for $T_{B}^{\ast}$ is a factor 0.36 smaller than the mean-field
prediction. We see that, in contrast to the mean-field result, the value of
$T_{p}$ obtained in the present treatment lies in the same range as the
experimentally determined temperature $T_{B}^{\ast}$. This coincidence is
worth remarking. However, the conclusions from the latter comparison of two
temperatures need care, because the temperature in \cite{Feld} is measured in
the weakly interacting regime and hence it may differ from the actual
temperature in the strongly interacting regime.

It is stated in Ref. \cite{Feld} that the discrepancy between the mean-field
and experimental pairing temperatures could suggest that the appearance of a
back-bending feature in the spectral function \cite{Gaebler}, which has been
interpreted as a signature for many-body pairing, is only a qualitative
evidence. However, the present results show that the fluctuations can
drastically reduce the pairing temperature $T_{p}$ with respect to
$T_{c}^{\ast}$. Thus there is no discrepancy between experiment and theory
when taking account the fluctuations.

\section{Conclusions \label{conclusions}}

The $T$-matrix approach straightforwardly applied to cold fermions in two
dimensions leads to a divergent fermion density for any finite temperature. We
have shown in the present work that taking into account the variation of the
order parameter in the number equations, as suggested in the GPF approach
\cite{Hu,Drum,Drum2}, provides a divergence-free description of the paired
state in two dimensions. This was shown both through numerical calculations
and through an analytic expansion at long wavelengths and low energies, where
the divergency occurs in the standard Nozi\`{e}res \& Schmitt-Rink approach.
The formalism allows to study the effects of the fluctuations both at zero and
at finite temperatures, and we find that fluctuations affect the critical
binding energy to obtain pairing and superfluidity in the presence of
imbalance. Moreover, the formalism also gives access to the density of states
of the pair fluctuations, from which we have defined a pseudogap temperature
$T_{p}$ as the temperature where an instability appears in the pair
fluctuation density. The pseudogap temperature defined in this way agrees with
the measured values of the pseudogap temperature in 2D fermi gases. Also the
location of the peaks in the spectral functions for pair fluctuations is shown
to agree with the experimental observations. The pseudogap temperature $T_{p}%
$, along with the critical temperature $T_{BKT}$ for superfluidity and the
pair formation temperature $T_{c}^{\ast}$, have been calculated as a function
of binding energy, temperature and imbalance, from which we obtain the phase
diagram as shown in Figs. 3 and 4. Whereas in mean-field the pseudogap
temperature is usually identified with the pair formation temperature, we find
that the inclusion of fluctuations beyond mean field strongly suppresses the
pseudogap temperature with respect to the mean-field pair formation
temperature. Moreover, in the presence of imbalance, the pseudogap temperature
may cross the BKT temperature for superfluidity. The results obtained here in
the context of superfluid quantum gases shed new light on the study of the
pseudogap phase in layered high-temperature superconductors, where the
question of the crossing of the pseudogap temperature with the superconducting
temperature, and the presence of preformed pairs, remains an open question.

\begin{acknowledgments}
Discussions with M. Zwierlein are gratefully acknowledged. This work was
supported by FWO-V projects G.0356.06, G.0370.09N, G.0180.09N, G.0365.08,
G.0115.12N, G.0119.12N, the WOG WO.033.09N (Belgium). J.~T. acknowlegdes
support of the Special Research Fund of the University of Antwerp under Grant
No. BOF NOI UA 2004.
\end{acknowledgments}


\begin{thebibliography}{99}                                                                                               %


\bibitem {DalibardReview}I. Bloch, J. Dalibard, and W. Zwerger, Rev. Mod.
Phys. \textbf{80}, 885 (2008).

\bibitem {deMelo1993}C. A. R. S\'{a} de Melo, M. Randeria, and J. R.
Engelbrecht, Phys. Rev. Lett. \textbf{71}, 3202 (1993).

\bibitem {MW}N. D. Mermin and H. Wagner, Phys. Rev. Lett. \textbf{17}, 1133 (1966).

\bibitem {Hohenberg}P. C. Hohenberg, Phys. Rev. \textbf{158}, 383 (1967).

\bibitem {PO}O. Penrose and L. Onsager, Phys. Rev. \textbf{104}, 576 (1956).

\bibitem {Yang}C.N. Yang, Rev. Mod. Phys. \textbf{34}, 694 (1962).

\bibitem {Kagan}Yu. Kagan, B.V. Svistunov and G.V. Shlyapnikov, Zh. Eksp.
Teor. Fiz. \textbf{93}, 552 (1987) [Sov. Phys. JETP \textbf{66}, 314 (1987)].

\bibitem {Berezinskii}V. L. Berezinskii, Sov. Phys. JETP \textbf{32}, 493 (1971).

\bibitem {KT}J. M. Kosterlitz and D. J. Thouless, J. Phys. C \textbf{6}, 1181
(1973); J. M. Kosterlitz, J. Phys. C \textbf{7}, 1046 (1974).

\bibitem {DalibardBKT}Z. Hadzibabic, P. Kr\"{u}ger, M. Cheneau, B. Battelier
and J. Dalibard, Nature \textbf{441}, 1118 (2006).

\bibitem {NSR}P. Nozi\`{e}res and S. Schmitt-Rink, J. Low Temp. Phys.
\textbf{59}, 195 (1985).

\bibitem {NSR2D}S. Schmitt-Rink, C. M. Varma, and A. E. Ruckenstein, Phys.
Rev. Lett. \textbf{63}, 445 (1989).

\bibitem {Hu}H. Hu, X.-J. Liu and P. D. Drummond, Europhys. Lett. \textbf{74},
574 (2006).

\bibitem {Drum}H. Hu, X. J. Liu, and P. D. Drummond, Phys. Rev. A \textbf{73},
023617 (2006).

\bibitem {Drum2}H. Hu, X. J. Liu, and P. D. Drummond, New Journal of Physics
\textbf{12}, 063038 (2010).

\bibitem {Salasnich}L. Salasnich and F. Toigo, J. Low Temp. Phys.
\textbf{165}, 239 (2011).

\bibitem {Feld}M. Feld, B. Fr\"{o}hlich, E. Vogt, M. Koschorreck, and M.
K\"{o}hl, Nature \textbf{480}, 75 (2011).

\bibitem {Sommer}A. T. Sommer, L. W. Cheuk, M. J. H. Ku, W. S. Bakr, and M. W.
Zwierlein, Phys. Rev. Lett. \textbf{108}, 045302 (2012).

\bibitem {VDM}D. van der Marel, Nat. Phys. \textbf{7}, 10 (2011).

\bibitem {Gaebler}J. P. Gaebler, J. T. Stewart, T. E. Drake, D. S. Jin, A.
Perali, P. Pieri, and G. C. Strinati, Nat. Phys. \textbf{6}, 569 (2010).

\bibitem {Perali}A. Perali, F. Palestini, P. Pieri, G. C. Strinati, J. T.
Stewart, J. P. Gaebler, T. E. Drake, and D. S. Jin, Phys. Rev. Lett.
\textbf{106}, 060402 (2011).

\bibitem {Palestrini}F. Palestini, A. Perali, P. Pieri, and G. C. Strinati,
Phys. Rev. B \textbf{85}, 024517 (2012).

\bibitem {Magierski}P. Magierski, G. Wlazlowski, and A. Bulgac, Phys. Rev.
Lett. \textbf{107}, 145304 (2011).

\bibitem {Tsuchiya}S. Tsuchiya, R. Watanabe, and Y. Ohashi, Phys. Rev. A
\textbf{80}, 033613 (2009).

\bibitem {LevinRPP}Q. Chen, J. Stajic, S. Tan, and K. Levin, Phys. Rep.
\textbf{412}, 1 (2005).

\bibitem {Babaev}E. Babaev and H. Kleinert, Phys. Rev. B \textbf{59}, 12083 (1999).

\bibitem {Gus1}V. P. Gusynin, V. M. Loktev, R. M. Quick, and S. G. Sharapov,
Int. J. Mod. Phys. B \textbf{12}, 3035 (1998).

\bibitem {Gus2}V. P. Gusynin, V. M. Loktev, and S. G. Sharapov, Zh. Eksp.
Teor. Fiz. \textbf{115}, 1243 (1999) [Sov. Phys. JETP \textbf{88}, 685 (1999)].

\bibitem {Traven}S. V. Traven, Phys. Rev. Lett. \textbf{73}, 3451 (1994).

\bibitem {Nelson1977}D. R. Nelson and J. M. Kosterlitz, Phys. Rev. Lett.
\textbf{39}, 1201 (1977).

\bibitem {Botelho2006}S. S. Botelho and C. A. R. S\'{a} de Melo, Phys. Rev.
Lett. \textbf{96}, 040404 (2006).

\bibitem {Popov}V. N. Popov, \emph{Functional Integrals in Quantum Field
Theory and Statistical Physics} (D. Reidel Publishing, Dordrecht, 1983).

\bibitem {BKT-PRA2009}J. Tempere, S. N. Klimin, and J. T. Devreese, Phys. Rev.
A \textbf{79}, 053637 (2009).

\bibitem {Stoof}J. O. Andersen, U. Al Khawaja, and H. T. C. Stoof, Phys. Rev.
Lett. \textbf{88}, 070407 (2002); U. Al Khawaja, J. O. Andersen, N. P.
Proukakis, and H. T. C Stoof, Phys. Rev. A \textbf{66}, 013615 (2002).

\bibitem {PS2002}N. Prokof'ev and B. Svistunov, Phys. Rev. A 66, 043608 (2002).

\bibitem {R1990}M. Randeria, J.-M. Duan, and L.-Y. Shieh, Phys. Rev. B
\textbf{41}, 327 (1990).

\bibitem {Tempere2007}J. Tempere, M. Wouters, and J. T. Devreese, Phys. Rev. B
\textbf{75}, 184526 (2007).

\bibitem {TKD2008a}J. Tempere, S. N. Klimin, J. T. Devreese, and V. V.
Moshchalkov, Phys. Rev. B \textbf{77}, 134502 (2008).

\bibitem {TKD2008b}J. Tempere, S. N. Klimin, and J. T. Devreese, Phys. Rev. A
\textbf{78}, 023626 (2008).

\bibitem {he}L. He and P. Zhuang, Phys. Rev. A \textbf{78}, 033613 (2008).

\bibitem {phasdiag3D}M. W. Zwierlein, A. Schirotzek, C. H. Schunck, and W.
Ketterle, Science \textbf{311}, 492 (2006); G. B. Partridge, W. Li, R. I.
Kamar, Y.-A. Liao, and R. G. Hulet, Science \textbf{311}, 503 (2006).

\bibitem {imbal3D}L. Radzihovsky and D.E. Sheehy, Rep. Prog. Phys. \textbf{73}
, 076501 (2010).

\bibitem {Petrov}D. S. Petrov and G. V. Shlyapnikov, Phys. Rev. A \textbf{64},
012706 (2001).

\bibitem {Michiel}M. Wouters, G. Orso, Phys. Rev. A \textbf{73}, 012707 (2006).

\bibitem {R1}R. B. Diener, R. Sensarma, and M. Randeria, Phys. Rev. A
\textbf{77}, 023626 (2008).

\bibitem {R2}R. B. Diener and M. Randeria, Phys. Rev. A \textbf{81}, 033608 (2010).

\bibitem {Keeling}J. Keeling, P. R. Eastham, M. H. Szymanska, and P. B.
Littlewood, Phys. Rev. B \textbf{72}, 115320 (2005).

\bibitem {Taylor2006}E. Taylor, A. Griffin, N. Fukushima, and Y. Ohashi, Phys.
Rev. A \textbf{74}, 063626 (2006).

\bibitem {OhashiPRA67}Y. Ohashi and A. Griffin, Phys. Rev. A \textbf{67},
063612 (2003).

\bibitem {Fukushima2007}N. Fukushima, Y. Ohashi, E. Taylor, and A. Griffin,
Phys. Rev. A \textbf{75}, 033609 (2007).

\bibitem {strinati}A. Perali, P. Pieri, L. Pisani, G.C. Strinati, Phys. Rev.
Lett. \textbf{92}, 220404 (2004).

\bibitem {strinati2}P. Pieri, L. Pisani, and G. C. Strinati, Phys. Rev. B
\textbf{72}, 012506 (2005).

\bibitem {KTD-JLTP2011}S. N. Klimin, J. Tempere and Jeroen P. A. Devreese,
Journal of Low Temperature Physics \textbf{165}, 261 (2011).

\bibitem {RanderiaNat}M. Randeria, Nature Physics \textbf{6}, 561 (2010).

\bibitem {QFS2010}S. N. Klimin and J. Tempere, Journal of Low Temperature
Physics \textbf{162}, 291 (2011).

\bibitem {Rourke}P. M. C. Rourke, I. Mouzopoulou, X. Xu, C. Panagopoulos, Y.
Wang, B. Vignolle, C. Proust, E. V. Kurganova, U. Zeitler, Y. Tanabe, T.
Adachi, Y. Koike, and N. E. Hussey, Nature Physics \textbf{7}, 455 (2011).
\end{thebibliography}
\end{document}